\documentclass[twocolumn,floatfix,superscriptaddress,amsmath,amsfonts,amssymb,aps]{revtex4}

\usepackage{graphicx}
\usepackage{dcolumn}
\usepackage{bm}
\usepackage[normalem]{ulem}
\usepackage{color}
\usepackage{amssymb}
\usepackage{amstext}

\begin{document}

\preprint{APS/123-QED}

\title{Collective magnetic response of inhomogeneous nanoisland FeNi films \\ around the percolation transition}

\author{N.~N. Kovaleva}
\affiliation{Lebedev Physical Institute RAS, 119991 Moscow, Russia}
\email{nkovaleva@sci.lebedev.ru}
\affiliation{Department of Physics, Loughborough University, LE11 3TU Leicestershire, United Kingdom}
\author{A.~V. Bagdinov}
\affiliation{Lebedev Physical Institute RAS, 119991 Moscow, Russia}
\author{\mbox{A. Stupakov}}
\author{A. Dejneka}
\affiliation{Institute of Physics ASCR, 18221 Prague, Czech Republic}
\author{E.~I. Demikhov}
\author{\mbox{A.~A. Gorbatsevich}}
\author{F.~A. Pudonin}
\affiliation{Lebedev Physical Institute RAS, 119991 Moscow, Russia}
\author{K. I. Kugel}
\affiliation{Institute for Theoretical and Applied Electrodynamics RAS, 125412 Moscow, Russia}
\author{F. V. Kusmartsev}
\affiliation{Department of Physics, Loughborough University, LE11 3TU Leicestershire, United Kingdom}

\date{\today}

\begin{abstract}
By using superconducting quantum interference device
(SQUID) magnetometry we investigated anisotropic high-field ($H$\,$\lesssim$\,7\,T) low-temperature (10\,K) magnetization response of inhomogeneous nanoisland FeNi films grown by rf sputtering deposition on Sitall (TiO$_2$) glass substrates. In the grown FeNi films, the FeNi layer nominal thickness varied from 0.6 to 2.5\,nm, across the percolation transition at the $d_c$\,$\simeq$\,1.8\,nm. We discovered that, beyond conventional spin-magnetism of Fe$_{21}$Ni$_{79}$ permalloy, the extracted out-of-plane magnetization response of the nanoisland FeNi films is not saturated in the range of investigated magnetic fields and exhibits paramagnetic-like behavior. We found that the anomalous out-of-plane magnetization response exhibits an escalating slope with increase in the nominal film thickness from 0.6 to 1.1\,nm, however, it decreases with further increase in the film thickness, and then practically vanishes on approaching the FeNi film percolation threshold.  At the same time, the in-plane response demonstrates saturation behavior above 1.5--2\,T, competing with anomalously large diamagnetic-like response, which becomes pronounced at high magnetic fields. 
It  is possible that the supported-metal interaction leads to the creation of a thin charge-transfer (CT) layer and a Schottky barrier at the FeNi film/Sitall (TiO$_2$) interface. Then, in the system with nanoscale circular domains, the observed anomalous paramagnetic-like magnetization response can be associated with a large orbital moment of the localized electrons. In addition, the inhomogeneous nanoisland FeNi films can possess spontaneous ordering of toroidal moments, which can be either of orbital or spin origin. The system with toroidal inhomogeneity can lead to anomalously strong diamagnetic-like response. The observed magnetization response is determined by the interplay between the paramagnetic- and diamagnetic-like contributions. 
\end{abstract}

\pacs{Valid PACS appear here}
\maketitle

Magnetic nanoparticles (NPs) are expected to play a decisive role in the development of next generation of ultra-high density magnetic data storage, quantum computing, magnetic targeted drug delivery, and many other applications (Sun and Murray 1999; Reiss and Hutten 2005; Frey and Sun 2011; McNally 2013; Mody et al. 2014).
On the other hand, magnetic NPs are attractive for fundamental research in magnetism. Magnetic properties of nanostructured materials, composed of metallic ferromagnetic (FM) NPs of small size, say 20\,--\,30 nm, or even smaller, essentially differ from those in the bulk having submicron size particles or larger. A small enough NP inevitably transforms into a single-domain FM  state, with parallel orientation of intraparticle atomic moments due to strong exchange interactions. The single-domain FM state of a NP can be associated with a  superspin (SS), comprising many atomic moments $\sim$\,10$^3$\,--\,10$^5$\,$\mu_{\rm B}$, where $\mu_{\rm B}$\,=\,9.27$\cdot$10$^{-{21}}$ emu is the Bohr magneton, the electron spin magnetic dipole moment. 
Weakly interacting NPs (small and well separated) exhibit the Curie-like behavior in a superparamagnetic (SPM) phase above the so-called blocking temperature ($T_B$), when the temperature is high enough to overcome the energy barrier between different orientations of a NP SS. However, when NPs are situated relatively close to each other, strong magnetic dipole-dipole interaction, exchange interaction (for sufficiently dense packing), and other nonclassical interactions, such as, for example, the Anderson-type superexchange coupling, originating from electron tunneling between the SPM NPs 
(Kondratyev and Lutz 1998), may become relevant. In particular, atomically small NPs, favoring the tunneling exchange between the neighboring bigger NPs, were considered as mediating the `superexchange' mechanism (Russier 2001; Kleemann et al. 2001). As a result, regular arrays of NPs, as well as their self-assembled local arrangements in a form of small clusters, can possess collective magnetic states at comparatively high temperatures.

Transitions into a superferromagnetic (SFM) phase were discovered in two-dimensional (2D) self-assembled (Kleemann et al. 2001; Rancourt and Daniels 1984; Miu et al. 2015; Stupakov et al. 2016) 
or regularly structured NP arrays (Russier 2001; Sugawara et al. 1997; Cowburn et al. 1999).
The collective states are characterised by the observed magnetic moment hysteretic-like irreversibility behaviour, induced by an external magnetizing field, which persists essentially above the blocking temperature $T_B$. The hysteretic-like contribution is the most pronounced at low temperatures, where  it exhibits a significantly slower relaxation rate. The origin of the SFM order in 2D granular systems is not well understood. Also, the relevance of two-dimensionality and additional `superexchange' interactions for the occurrence of the out-of-plane SFM behavior in quasi-2D FM NPs arrays remains unclear.

Yet another specific issue, associated with a domain state, has not been significantly explored in the physics of 2D granular systems. Nevertheless, the relevant physics was elaborated for 2D layers of diamagnetic organic molecules forming self-organized domains on gold. It was demonstrated that for a circular domain, the lowest energy modes, corresponding to rotations around the domain axis, with a high angular momentum value may exist (Vager and Naaman 2004a). 
In addition, the combination of spin-orbit coupling and contact potential at the large radius domain boundaries may lead to a `giant' orbital moment induced in atomiclike localized electronic states (Hernando 2006).
In agreement with these theoretical predictions, large FM-like magnetization, up to several tens of Bohr magnetons per adsorbed organic molecule, with a very small hysteresis and no saturation up to a field of 1 T, with the highest response along the axis perpendicular to the surface, was observed (Carmeli et al. 2003; Vager et al. 2004b). 

Unfortunately, high-field magnetic response of FM nanoscaled domain systems has been only scarcely investigated. For example, it is reported that nanocomposite systems, 
combined of nanostructured silicon and embedded metallic Ni nanostructures, show low-field FM spin-magnetism of the incorporated Ni nanostructures and an additional non-saturating up to 7\,T para-magnetic-like term (Rumph 2008, 2010).
It is suggested that the non-saturating paramagnetic-like magnetization 
term arises due to orbital mesoscopic persistent currents driven by the 
symmetry breaking at the metallic nanostructure\,--\,semiconducting silicon interface due to the Rashba field (Ganichev et al. 2013). However, the origin of orbital currents in FM nanoscaled domain systems needs to be comprehensively investigated.    
 
Here, by using SQUID magnetometry, we study  high-field (up to 7\,T) low-temperature ($T$\,$\simeq$\,10\,K) magnetization response of the inhomogeneous nanoisland FeNi film samples with the nominal film thickness varying from 0.6 to 2.5 nm, across the physical percolation threshold at the $d_c$\,$\simeq$\,1.8\,nm (Stupakov et al. 2016; Sherstnev 2014; Boltaev 2017),
in the planar and perpendicular geometry of an applied dc magnetic field. The nanoisland FeNi films were grown by rf sputtering deposition onto Sitall glass substrates. The Sitall material, utilized in our experiments, is represented by TiO$_2$ rutile phase. 
Beyond conventional spin-magnetism of Fe$_{21}$Ni$_{79}$ permalloy, which is known to saturate at low magnetic fields, we discovered that the out-of-plane magnetization response of the discontinuous FeNi films exhibits paramagnetic-like behavior in the range of investigated magnetic fields up to 7\,T. The anomalous out-of-plane magnetization response quickly decreases on approaching the film percolation threshold due to the nanoisland coalescence. For the same nominal thickness range of the inhomogeneous nanoisland FeNi films on the Sitall (TiO$_2$) substrate, the out-of-plane SFM magnetization behavior was found in our recent study 
(Stupakov et al. 2016). Moreover, the high-field magnetization response exhibits remarkable anisotropy for the studied nanoisland FeNi film samples, where the in-plane response is significantly suppressed. There is a trend that the diamagnetic-like response, competing with the FM-like behavior, becomes clearly recognizable above 1.5--2\,T in the in-plane response. It is interesting that for the FeNi film with the thickness slightly above the FeNi film percolation threshold, the in-plane diamagnetic-like magnetization response was observed for the whole range of investigated dc magnetic fields. Here, we also compare the observed magnetization response with the in-plane magnetization response of the FeNi films sputtered onto crystalline diamagnetic substrates Al$_2$O$_3$ (100). The origin of the `exotic' magnetic phases, discovered in the inhomogeneous FeNi nanoilsand layers, grown onto the Sitall (TiO$_2$) glass substrates, and identified by their anomalous paramagnetic-  and diamagnetic-like response, is discussed. The associated high-field magnetic properties are important from the practical point of view and can be utilized, for example, in high magnetic field sensing devices.\\

\begin{figure}
  \includegraphics[width=0.47\textwidth]{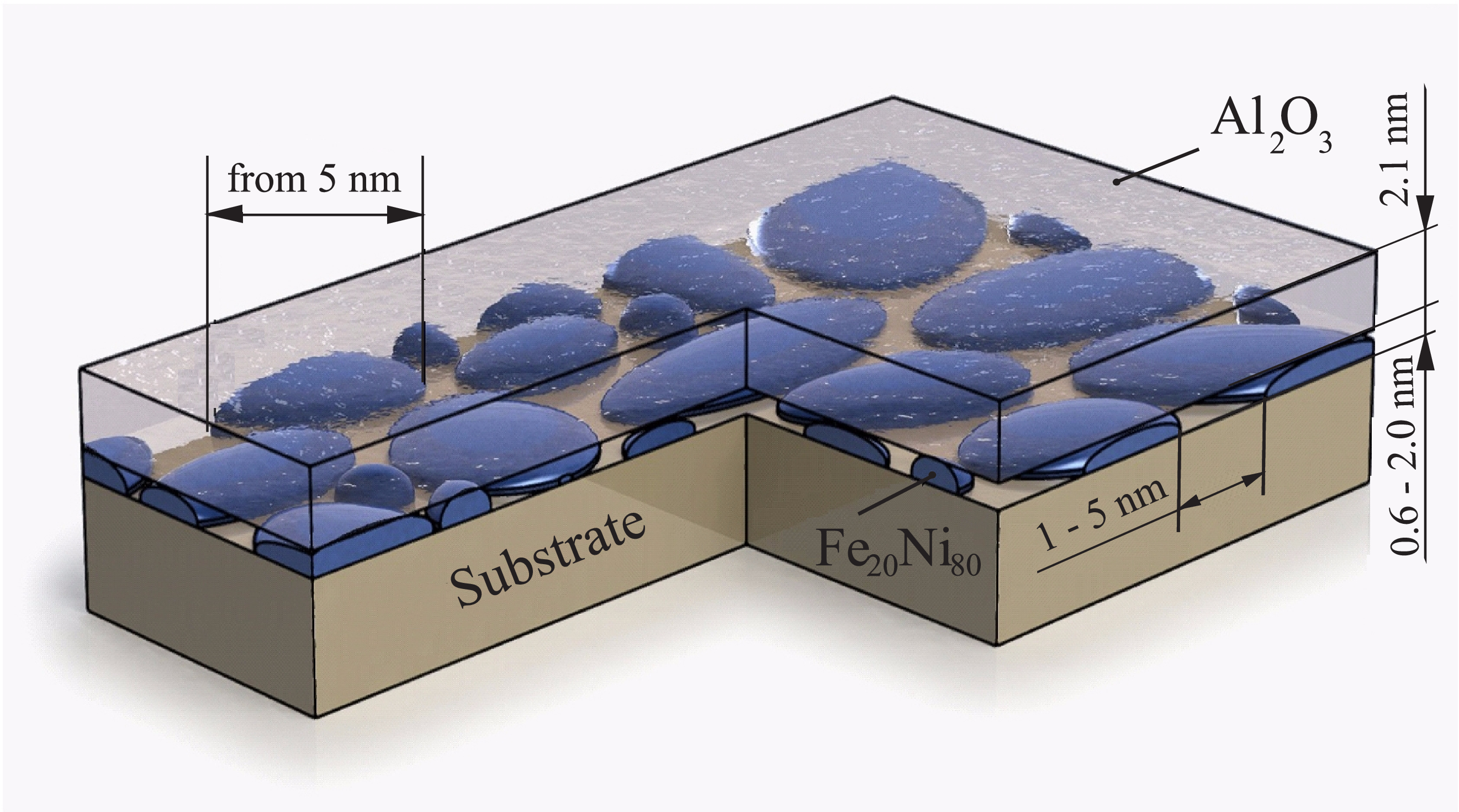}
\caption{A schematic picture of the nanoisland FeNi film samples (capping Al$_2$O$_3$ layer (2.1 nm)/FeNi($d$)/Sitall substrate).}\vspace{-1.2em}
\label{Schem}       
\end{figure}

\hspace{-1.0em}{\bf FeNi film samples}. The nanoisland FeNi films were grown by rf sputtering deposition from 99.95\% pure Fe$_{21}$Ni$_{79}$ targets at a base vacuum pressure less than 2$\times$10$^{-6}$ Torr and a background Ar pressure of 6$\times$10$^{-4}$ Torr. We used glass-ceramic Sitall substrates and crystalline Al$_2$O$_3$(100) substrates. To reduce the influence of a surface crystalline anisotropy, an amorphous 7\,nm thick Al$_2$O$_3$ buffer layer was deposited on top of the Al$_2$O$_3$ substrates. An analysis of the X-ray diffraction pattern of the Sitall substrate, used in the present experiments, showed that it contains the crystalline TiO$_2$ rutile phase (Kovaleva et al. 2016). 
In contrast to the Al$_2$O$_3$ substrate, the used glass-ceramic Sitall substrate provides good wetting conditions for the FeNi adhesion (this issue is considered in detail in the Discussion section). Our spectroscopic ellipsometry study of the FeNi films of different thickness grown on the Sitall substrates demonstrated that their dielectric permittivity changes from insulating- to metallic-like at the critical film thickness of about 1.8 nm. In addition, the temperature dependence of their dc conductivity suggests the existence of a percolation transition for this critical FeNi film thickness (Sherstnev 2014; Boltaev et al. 2017). 
Below the percolation transition at the critical thickness $d_c$\,$\simeq$\,1.8 nm, the FeNi layer has a discontinuous structure (Stupakov et al. 2016),
where lateral sizes of the nanoislands are 5--30 nm, and the distance between them is less than 5 nm (as schematically illustrated by Fig.\,\ref{Schem}). For the present study, we prepared a series of the FeNi films grown on the Sitall substrates, with the nominal film thickness varying from 0.6 to 2.5 nm. The nominal film thickness (that is, the thickness of the corresponding continuous film) was determined by the deposition time defined by the film deposition rate ($\sim$\,0.67 \AA/sec). During the deposition, the substrate temperature was 73\,$\pm$\,3\,$^\circ$C only. To avoid oxidation of the films at ambient conditions, the grown FeNi films were covered {\it in situ} by a 2.1\,nm thick Al$_2$O$_3$ capping layer (as schematically illustrated by Fig.\,\ref{Schem}).\\

\begin{figure}
\hspace{0.45cm} \includegraphics[width=0.375\textwidth]{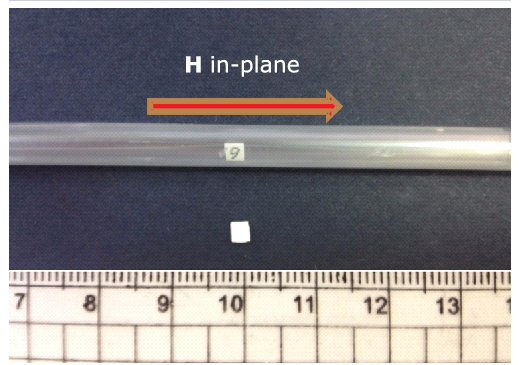}
\caption{(a) The FeNi film sample (capping Al$_2$O$_3$ layer (2.1 nm)/FeNi film/Sitall substrate), mounted for the SQUID measurements in the in-plane $H$-field geometry.} 
\vspace{-1.4em}
\label{Fig1}   
\end{figure}

\hspace{-1.0em}{\bf Magnetization measurements}. For magnetization measurements we cut out the samples of approximate dimensions 3$\times$3 mm$^2$ (as it is illustrated by Fig.\,\ref{Fig1}). Below we give the details of the FeNi film magnetization evaluation. Once the total magnetization $M_0$ of the FeNi film sample of the mass $m_{0}$, as well as the magnetization $M_{s0}$ of the blank substrate of the mass $m_{s0}$, are measured, the FeNi film magnetization $M_f$ can be evaluated from the following simple considerations. The total sample mass $m_0$ can be expressed as 
$m_0=m_s+m_f=\rho_sSL+\rho_fSd$,
where $m_s$ is the substrate mass, 
$m_f$ is the film mass, $S$ is the sample surface area, $L$ is the substrate thickness, $d$ is the nominal film thickness, $\rho_{f}$ is the film density, and  
$\rho_s$ is the substrate density.
The total film sample magnetization $M_0=M_s+M_f$ 
is  the sum of contributions of the substrate and film magnetization, $M_s$ and $M_f$, respectively. In turn, the substrate contribution, $M_s$, can be evaluated from the magnetization $M_{s0}$, obtained from the independent magnetization measurements on the blank substrate sample of the mass $m_{s0}$ as 
$M_s=\frac{M_{s0}\cdot m_s}{m_{s0}}$.
Then, the total film magnetization $M_f$ can be evaluated via the FeNi film sample mass $m_0$ as
$M_f=M_0-M_s=M_0-\frac{M_{s0} \cdot m_s}{m_{s0}}=
M_0-\frac{M_{s0}\cdot \left( m_0-m_f \right)}{m_{s0}}
\cong\ \left[m_f \ll\ m_0 \right] \cong\ M_0-\frac{M_{s0} \cdot m_0}{m_{s0}}$. 
However, to compare the FeNi film magnetization values $M_f$ measured on a series of the film samples, having different mass and size, one needs the normalized magnetization values. 
Since 
$S=\frac{m_0}{\rho_sL+\rho_fd}\cong \left[ d\sim 10^{-9}\ {\rm m},\ L\sim10^{-3}\ {\rm m} \right] \cong \frac{m_0}{\rho_s L}$,
we arrive at the following formula for the evaluation of the FeNi film magnetization normalized to the sample surface area $S$ 
\begin{equation}
\frac{M_f}{S}\cong \frac{\rho_s L}{m_0}\left( M_0-\frac{M_{s0}\cdot m_0}{m_{s0}}\right)=\rho_s L\left( \frac{M_0}{m_0}-\frac{M_{s0}}{m_{s0}}\right)
\label{Eq1}
\end{equation}
expressed via the substrate thickness $L$ and the substrate density $\rho_s$.   
We note that the FeNi films investigated in the present study were sputtered onto the Sitall substrates cut from a single Sitall material sheet, having the thickness $L$\,$\simeq$\,0.056 $\pm$ 0.004 cm, where $\frac{\Delta L}{L}\simeq$ 7\%. The estimated Sitall material density is $\rho_s$\,$\simeq$\,2.72 ($\pm\,4\%$) g/cm$^3$. The additional errors in the use of Eq.\,(\ref{Eq1}) come from the accuracy of the measured quantities $\Delta M_0$ and $\Delta M_{so}$ (of about 2\,$\cdot$\,10$^{-8}$ emu) and $\Delta m_0$ and $\Delta m_{so}$ (of about 0.04 mg). Typical values of the measuring quantities $M_0$ and $M_{s0}$ (at $T\simeq$\,10 K and $H\simeq$\,6.5\,T) are in the range 0.001\,--\,0.005 emu for the studied FeNi sample series, so $\frac{\Delta M_0}{M_0}\lesssim$ 0.002\%. Typical values of the $m_0$ and $m_{s0}$ are about 10 mg, so $\frac{\Delta m_0}{m_0}\simeq$ 0.4\%. Thus, in the magnetization evaluation using Eq.\,(\ref{Eq1}), for $\frac{M_0}{m_0}$ we have $\frac{\Delta \left( \frac{M_0}{m_0} \right)}{\frac{M_0}{m_0}}=\frac{\Delta M_0}{M_0}+\frac{\Delta m_0}{m_0}\simeq0.4\%$, and, ultimately, $\frac{\Delta \left( \frac{M_f}{S} \right)}{\frac{M_f}{S}}=\frac{\Delta \rho_s}{\rho_s}+\frac{\Delta L}{L}+\frac{\Delta \left( \frac{M_0}{m_0} \right)}{\frac{M_0}{m_0}} \simeq11.5\%$. Evaluating the FeNi film sample magnetization using Eq.\,(\ref{Eq1}), normalized to the sample surface area rather than to the film volume, let us to avoid a source of large potential errors, associated with uncertainty in the nominal film thickness.
\begin{figure*}
  \includegraphics[width=0.99\textwidth]{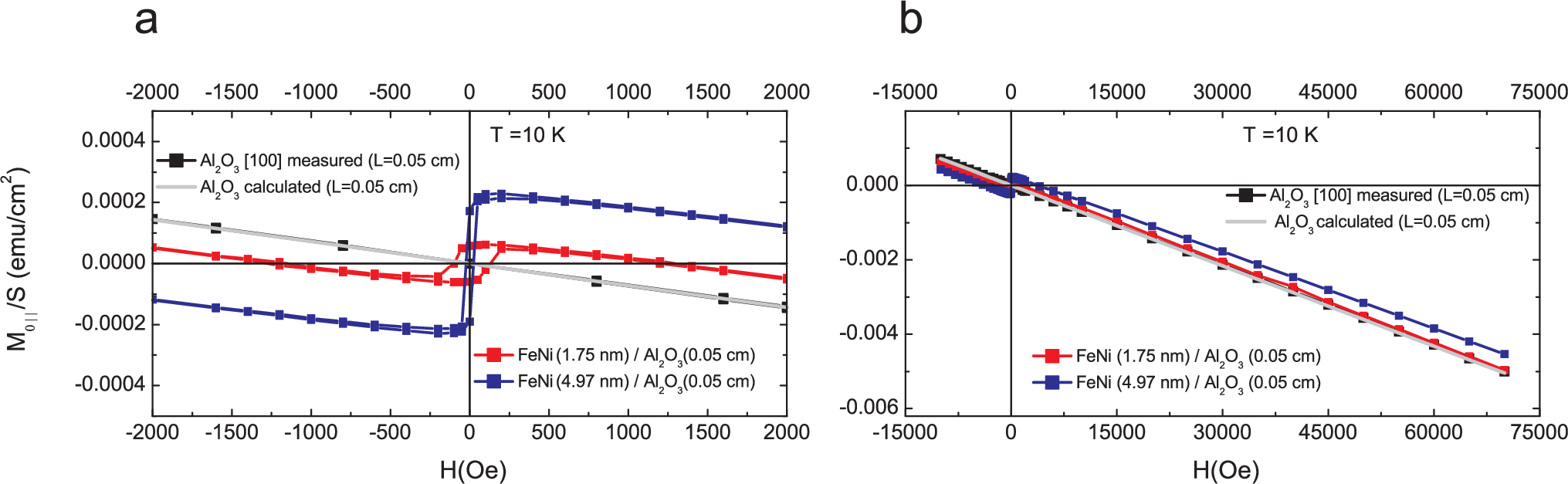}
\caption{In-plane magnetization response of the FeNi films on the Al$_2$O$_3$ substrate. (a) Low-field and (b) high-field magnetization response (in emu/cm$^2$) of the FeNi film samples -- capping Al$_2$O$_3$ layer(2.1\,nm)/FeNi film ($d$\,$\simeq$\,1.75 and 4.97\,nm)/buffer Al$_2$O$_3$ layer (7.0\,nm)/Al$_2$O$_3$  (100) substrate (0.05\,cm). The displayed symbols are larger than the error bars. Notice good agreement between the measured and calculated magnetization response for the blank Al$_2$O$_3$ substrate in (a,b).}\vspace{-1.2em}
\label{DiaSubs}       
\end{figure*}

\begin{figure}
  \includegraphics[width=0.47\textwidth]{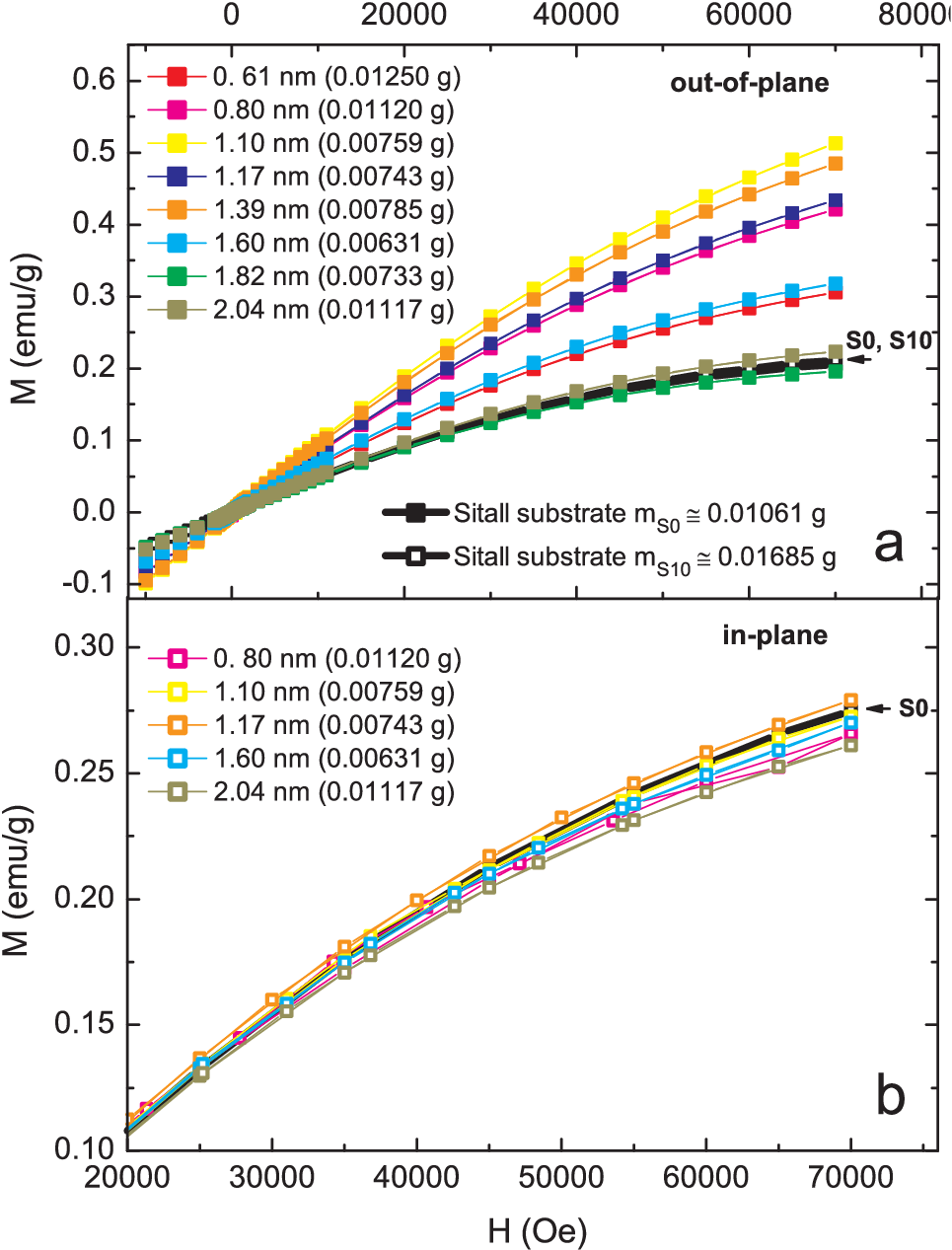}
\caption{The mass-normalized magnetization response of the FeNi film samples (the sample masses are given in brackets) and of the blank Sitall substrate samples (S0 and S10) in the magnetic field applied  in zero-field-cooling conditions in the (a) out-of-plane and (b) in-plane geometry. The displayed symbols are larger than the error bars. Additional errors determined by assignment of the demagnetization factor  due to possible sample misalignment are not large (notice good coincidence of the mass-normalized data for the blank substrate samples). The solid curves are the guides-to-the-eye.}\vspace{0.5 cm}
\label{Mtom}       
\end{figure}

Here, by using a Quantum Design SQUID magnetometer (MPMS XL 7\,T), we measured magnetization of the grown FeNi film samples at $T$\,$\simeq$\,10 K for the dc magnetic fields --\,1.5\,T\,$\lesssim H \lesssim$\,7\,T, applied along an in-plane sample direction (as it is illustrated by Fig.\,\ref{Fig1}) and perpendicular to the sample surface. High sensitivity of magnetic measurements (2\,$\times$\,10$^{-8}$ emu) was enabled by reciprocating sample transport. Recently by using a MPMS XL 7\,T SQUID magnetometer total magnetization of the $d^0$ charge-imbalanced  ferromagnetic interface between nonmagnetic perovskites, such as  SrTiO$_3$ / KTaO$_3$, SrTiO$_3$ / KNbO$_3$, and SrTiO$_3$ / NaNbO$_3$, of the order of 10$^{-6}$ emu was reported (Oja et al. 2012).
The FeNi film magnetization in the present study was estimated in accordance with Eq.\,(\ref{Eq1}), namely, by subtracting the mass-normalized magnetization of the blank Sitall substrate from the mass-normalized magnetization response of the FeNi film sample (capping Al$_2$O$_3$ layer (2.1 nm)/FeNi film\,($d$)/Sitall substrate). In this case, an accuracy of the FeNi film magnetization estimate will be also determined by the accuracy of the sample alignment in the out-of-plane or in-plane configuration. Nevertheless, we argue that the accuracy of alignment setting in our magnetometer was appropriate, as, for example, the magnetization data obtained for different samples of the blank Sitall substrate in out-of-plane $H$-field configuration coincide with a high accuracy (see Fig.\,\ref{Mtom}(a)).\\

\hspace{-1.0em}{\bf Low- and high-field magnetization response of the FeNi films on the Al$_2$O$_3$ substrates}.
Figure\,\ref{DiaSubs}(a,b) shows the in-plane low-field ($-0.2$\,T\,$\lesssim$\,$H$\,$\lesssim$
\,0.2\,T) and high-field ($-1.5$\,T\,$\lesssim$\,$H$\,$\lesssim$\,7.0\,T) magnetization response ($T$\,$\simeq$\,10 K) of the 1.75 and 4.97\,nm thick FeNi film samples [capping Al$_2$O$_3$ layer (2.1 nm)/FeNi film ($d$)/buffer Al$_2$O$_3$ layer (7.0 nm)/Al$_2$O$_3$  substrate]. The presented magnetization response (in emu/cm$^2$) is estimated using the simple relationship  $\frac{M_{0\parallel}}{S}=\rho L\frac{M_{0\parallel}}{m_0}$, where $M_{0\parallel}$ is the measured in-plane film sample magnetization, $m_0$ is the film sample mass, $\rho$\,$\simeq$\,3.97 g/cm$^3$ is the density of bulk Al$_2$O$_3$, and $L\simeq$\,0.05\,cm is the Al$_2$O$_3$(100) substrate thickness. The Al$_2$O$_3$ (100) substrate contribution was obtained independently from SQUID measurements on the blank substrate sample for the dc magnetic fields $-1.5$\,T\,$\lesssim$\,$H$\,$\lesssim$\,7.0\,T in the in-plane $H$-field geometry. From Fig.\,\ref{DiaSubs}(a,b) one notices that the Al$_2$O$_3$ substrate sample exhibits the diamagnetic magnetization field dependence. We obtained good agreement between the measured and calculated values for the magnetization field dependence of the Al$_2$O$_3$ substrate sample, using the molar susceptibility $\chi_m$\,$\simeq$\,
$-$37\,$\cdot$\,10$^{-6}$ cm$^3$/mol (see Fig.\,\ref{DiaSubs}(b)). In the low-field magnetization response of the FeNi film samples shown in Fig.\,\ref{DiaSubs}(a) one can clearly see the magnetization loops around zero field, which can be identified with the FM FeNi layer. Also, one notices that the 1.75 and 4.97\,nm thick FeNi films show the saturation magnetization behavior on a background of the Al$_2$O$_3$ substrate diamagnetic response. For 78.5\% permalloy, the saturation of intrinsic induction $B$\,--\,$H$\,$\simeq$\,10800 G, corresponding to the saturation magnetization $M_s$\,$\simeq$\,860\,G, occurs for the magnetizing force $H$\,$\gtrsim$\,10\,Oe\,$\simeq$\,10$^{-3}$\,T (Elmen 1935). 
Then, the saturation magnetization of the 1\,nm thick Fe$_{21}$Ni$_{79}$ permalloy layer for a typical film sample with the surface area $S$=10$^{-1}$ cm$^2$  can be estimated as $\sim$\,8.6\,$\cdot$\,10$^{-6}$ emu. It is known that submicron permalloy dot arrays demonstrate the saturation magnetization peculiar of continuous permalloy films (Schneider and Hoffmann 1999).
From Fig.\,\ref{DiaSubs}(a) we estimate that the saturation magnetization of the FeNi film samples (normalized to the nominal FeNi film thicknesses of 1.75 and 4.97\,nm) is $\sim$\,5.4\,$\cdot$\,10$^{-6}$ and $\sim$\,5.3\,$\cdot$\,10$^{-6}$ emu, respectively. The agreement is good enough for these ultrathin FeNi films, however, we note that the estimated values are somewhat smaller. It is also known that thin film permalloy samples with the thickness until 100 nm show saturation magnetization at the applied magnetic field $H_{as}$\,$\simeq$\,5\,Oe (Kern et al. 2016). 
Here we observe that that the magnetization loop for the 4.97 nm thick FeNi film is rather narrow (see Fig.\,\ref{DiaSubs}(a)), indicating that this film is nearly continuous. However, the saturation field for the submicron dot arrays, for example, for magnetically saturated oblate permalloy ellipsoids (with diameter $d$ and height $h$), which is determined by the shape anisotropy given by the aspect ratio, $r=d/h$, is different. Approaching saturation magnetization, the applied field of saturation $H_{as}$ has to equal to the intrinsic demagnetizing field, $H_d$\,=\,$-NM_s$, where N is the demagnetization factor, which is determined by the demagnetizing field of isolated, magnetically saturated dot. Here, we observe that the saturation magnetization for the nanoiland 1.75 nm thick FeNi film occurs at the applied magnetic field $H_{as}$\,$\simeq$\,250\,--\,300\,Oe (see Fig.\,\ref{DiaSubs}(a)). From the dependence of the demagnetizing field as function of the aspect ratio (Schneider and Hoffmann 1999), 
the in-plane demagnetizing field 250\,--\,300\,Oe corresponds to the aspect ratio of 25, which gives an estimate for the  average circular base diameter in the nanoiland 1.75 nm thick FeNi film of $\sim$\,40\,nm. 

Thus, the studied in-plane low-field magnetization response of the FeNi films grown onto the Al$_2$O$_3$ substrates indicates that the 1.75 nm thick FeNi film has a nanoisland structure, whereas the 4.97 nm thick FeNi film, seemingly, has a nearly continuous structure. In the studied high-field magnetization response, these films show magnetization saturation behavior, peculiar of per-malloy films. In general, no anomalies were discovered, and the micromagnetics of the FeNi films on the Al$_2$O$_3$ substrates can be satisfactorily described by the film's morphology.\\

\hspace{-1.0em}{\bf High-field magnetization response of the nanoisland FeNi films on the Sitall substrates}.
Figure\,\ref{Mtom}(a,b) presents mass-normalized magnetization response of the grown nanoisland FeNi film samples\,[cap-ping Al$_2$O$_3$ layer (2.1\,nm)/FeNi film ($d$)/Sitall substrate] with the nominal film thickness $d$ varying from 0.61 to 2.0\,nm. The magnetization response was measured at $T$\,$\simeq$\,10 K for the dc magnetic fields --\,1.0\,T\,$\lesssim$\,$H$\,$\lesssim$\,7\,T applied along an in-plane sample  direction and perpendicular to the sample surface. Focusing at high magnetic fields in Fig.\,\ref{Mtom}(a), one can clearly follow the evolution of the out-of-plane mass-normalized magnetization response, which first increases and subsequently decreases with increasing the film thickness from 0.61 to 1.82\,nm. By contrast, variation of the in-plane mass-normalized magnetization response is comparatively much less pro-nounced in the studied range of thicknesses (see Fig.\,\ref{Mtom}(b)).

\begin{figure}
  \includegraphics[width=0.46\textwidth]{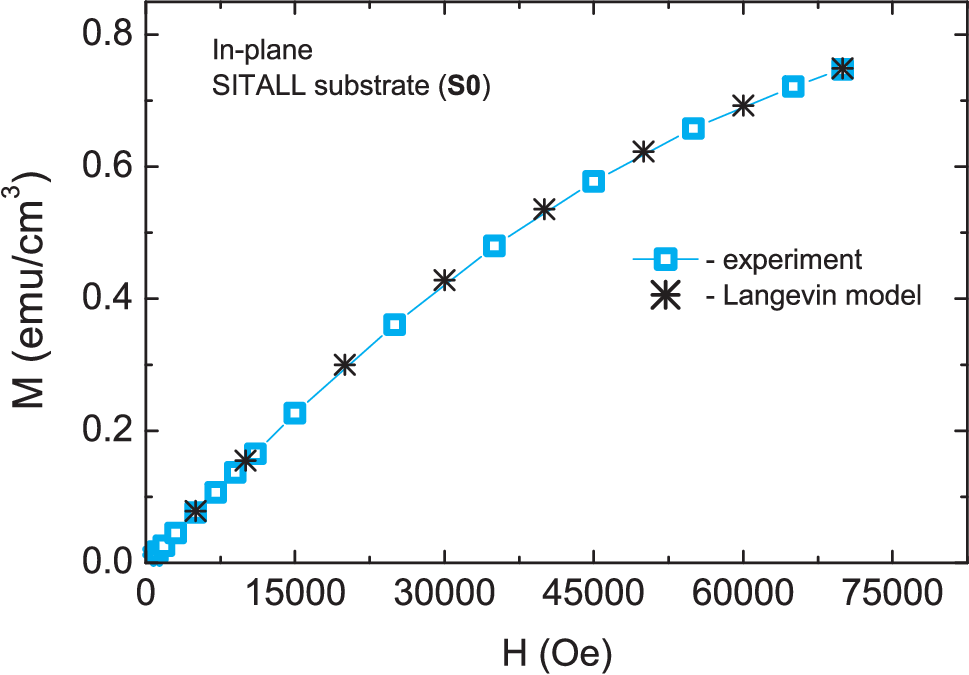}
\caption{The in-plane magnetization field dependence of the blank Sitall substrate and the fit result with the Langevin function, $M(H,T)$\,=\,$N\mu_p\left[ {\rm coth}\left( \frac{\mu_p H}{k_{\rm B}T} \right)-\frac{k_{\rm B}T}{\mu_pH} \right]$. The symbols are less than the error bars.}\vspace{-1.2em}
\label{Sit_Lang}       
\end{figure}

\begin{figure}
  \includegraphics[width=0.47\textwidth]{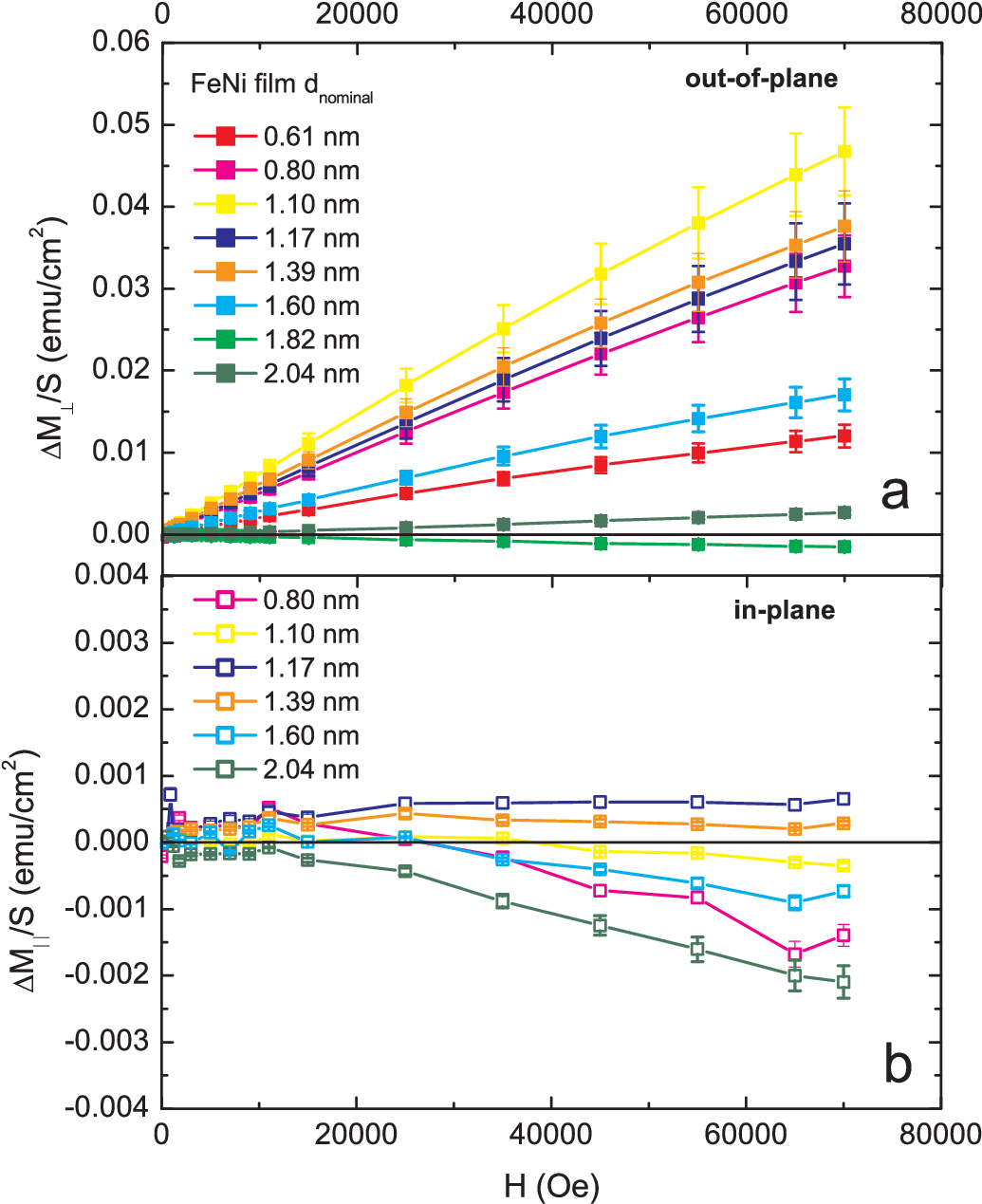}
\caption{The effective (a) out-of-plane, $\Delta M_{\perp}(H)/S$, and (b) in-plane, $\Delta M_{\parallel}(H)/S$, magnetization response of the nanoisland FeNi films ($T$\,$\simeq$\,10 K), obtained from Eq.\,(\ref{Eq1}). The solid lines are the guides-to-the-eye.}\vspace{-1.2em}
\label{DeltaM}       
\end{figure}

To extract the effective FeNi film magnetization, the substrate contribution was obtained independently from SQUID measurements on the blank Sitall substrate sample at $T$\,$\simeq$\,10 K for the dc magnetic fields \mbox{--\,1.0\,T}\,$\lesssim$\,$H$\,$\lesssim$\,7\,T in the corresponding $H$-field geometries (see Fig.\,\ref{Mtom}(a,b)). The substrate sample showed unsaturated, Langevin-like, magnetization field dependence, which can be associated with the presence of magnetic impurities and/or defects pertaining the used Sitall glass material (Kovaleva et al. 2016). 
We fitted the in-plane substrate magnetization with the Langevin function,\\ $M(H,T)$\,=\,$N\mu_p\left[ {\rm coth}\left( \frac{\mu_p H}{k_{\rm B}T} \right)-\frac{k_{\rm B}T}{\mu_pH} \right]$, where $k_B$ is the Boltzman's constant (see Fig.\,\ref{Sit_Lang}), and estimated the average magnetic moment of magnetic impurities $\mu_p$\,$\simeq$\,6\,$\pm$
\,0.02\,$\mu_B$ and their concentration $N_p$\,$\simeq$\,(2.03\,$\pm$\,0.02)\,$\cdot$10$^{19}$ cm$^{-3}$. We note that the diamagnetic contribution from the capping Al$_2$O$_3$ layer (of 2.1\,nm thick, sputtered on the sample surface area of about $S$=10$^{-1}$ cm$^2$) is beyond the sensitivity of the present SQUID measurements. Indeed, taking the molar susceptibility $\chi_m$ and the density $\rho$ of bulk Al$_2$O$_3$, we estimated its diamagnetic contribution at 7\,T of about $-$\,2\,$\cdot$\,10$^{-9}$ emu.      
The effective FeNi film magnetization, $M_f$, normalized to the film surface area $S$, was estimated according to Eq.(1).
Estimating the effective FeNi film magnetization normalized to a sample surface area (rather than to a film volume) let us avoiding the source of additional potential errors due to the uncertainty in the nominal FeNi film thickness. By subtracting, in accordance with Eq.\,(1), the mass-normalized magnetization of the Sitall substrate sample from the mass-normalized magnetization of the FeNi film samples, measured at the dc magnetic fileds --\,1\,T\,$\lesssim$\,$H$\,$\lesssim$\,7\,T in the corresponding $H$-field geometries at $T$\,$\simeq$\,10 K (Fig. \ref{Mtom}(a,b)), we obtained the effective out-of-plane, $M_{f \perp}(H)/S$, and in-plane, $M_{f \parallel}(H)/S$, magnetization of the studied nanoisland FeNi films (shown in Fig.\,\ref{DeltaM}(a,b), respectively).    

Now we analyze behavior of the extracted effective magnetization, $M_{f \perp}(H)/S$ and $M_{f \parallel}(H)/S$, as function of the nominal FeNi film thickness. One notices that for the FeNi films with the nominal thickness varying from 0.61 to 1.6 nm the $M_{f \perp}(H)/S$ is not saturated in the range of investigated magnetic fields and exhibits nearly linear with $H$ magnetization response (see Fig.\,\ref{DeltaM}(a)). One can follow that the $M_{f \perp}(H)/S$ exhibits an escalating slope with increase in the nominal film thickness from 0.61 to 1.1 nm. Here, the slope of the $M_{f \perp}(H)/S$ seemingly correlates with an average size of the FeNi nanoislands, which increases with nominal film thickness. Apparently, the maximum of the $M_{f \perp}(H)/S$ is obtained for the FeNi film thickness, where the nanoislands have the largest average size, but are still definitely separated from each other. With increasing film thickness and approaching the FeNi film percolation threshold due to the nanoisland coalescence at the $d_c$\,$\simeq$\,1.8 nm,
the slope of the $M_{f \perp}(H)/S$ quickly decreases. We would like to mention that the observed out-of-plane high-field unsaturated behavior corresponds to the out-of-plane SFM magnetization behavior for the nanoiland FeNi films on the Sitall substrate with the nominal film thickness 0.6\,nm\,$\lesssim$\,$d$\,$\lesssim$\,1.8\,nm, found in our recent study (Stupakov et al. 2016).

The effective magnetization shown in Fig.\,\ref{DeltaM}(a,b) exhibits remarkable anisotropy for the nanoisland FeNi films with the thicknesses in the 0.8\,nm\,
$\lesssim$\,$d$\,$\lesssim$\,1.6\,nm range, where the in-plane response is significantly suppressed. One notices the following trends in the in-plane effective magnetization field dependence, $M_{f \parallel}(H)/S$, as function of the nominal film thickness (Fig.\,\ref{DeltaM}(b)). Thus, FM-like saturation behavior, with the saturated magnetization value $M_{f \parallel}(H)/S$\,$\simeq$
6$\cdot$10$^{-4}$\,emu/cm$^2$ above 1.5\,--\,2\,T, was observed for the FeNi film with the nominal thickness 1.17 nm (we note that this thickness is slightly above 1.10\,nm, for which the $M_{f \perp}(H)/S$ maximum was revealed). The saturation behavior observed above 1.5\,--\,2\,T cannot be associated with the permalloy layer. For example, we observed that the saturation magnetization for the nanoiland 1.75 nm thick FeNi film grown on the Al$_2$O$_3$ substrate occurs at the applied magnetic field $H_{as}$\,$\simeq$\,250\,--\,300\,Oe (see Fig.\,\ref{DiaSubs}(a)). In addition, taking the typical sample area $S$=10$^{-1}$ cm$^2$, the effective saturation magnetization response, normalized to the film thickness can be estimated as 5\,$\cdot$\,10$^{-5}$ emu, which is 6 times larger than the saturation magnetization  of the 1 nm thick Fe$_{21}$Ni$_{79}$ permalloy layer (estimated to be of about 8.6\,$\cdot$
\,10$^{-6}$ emu, as we already mentioned above). Also, this effect cannot be attributed to the magnetization induced in the near-interface region. We estimated that magnetization induced by dipole-dipole interaction of a SS magnetic moment of FM FeNi nanoislands and SPM impurities of the Sitall substrate (with $\mu_p$\,$\simeq$\,6\,$\mu_B$ and their concentration $N_p$\,$\simeq$\,2.03\,$\cdot$10$^{19}$ cm$^{-3}$) can hardly contribute here essentially (with the effect of $\sim$\,0.006\% only). 

In addition, there is a trend that diamagnetic-like response, competing with the FM-like response for the FeNi film samples with the nominal thickness of 0.8, 1.1, and 1.6\,nm, becomes clearly recognizable above 1.5\,--\,2\,T at high $H$ (see Fig.\,\ref{DeltaM}(b)). Moreover, for the FeNi film with the thickness 2.0\,nm, slightly above the FeNi film percolation threshold due to the nanoisland coalescence at the $d_c$\,$\simeq$\,1.8 nm, the in-plane diamagnetic-like magnetization response, $M_{f \parallel}(H)/S$, was observed for the whole  range of the studied dc magnetic fields (see Fig.\,\ref{DeltaM}(b)). For the evaluation of the effective diamagnetic volume susceptibility, one should substitute the effective thickness of the layer responsible for the anomalous diamagnetic-like response. According to our atomic-force microscopy (AFM) study of the grown FeNi films (Stupakov et al. 2016), a large-scale topography profile of the used Sitall substrate indicates the height variation in the range 1\,--\,3\,nm, which characterizes its surface roughness. The effective thickness of the layer responsible for the anomalous diamagnetic response should be estimated taking into account the substrate surface roughness. More issues regarding the evaluation of the effective diamagnetic susceptibility are addressed in the Discussion section.  

\begin{figure}
  \includegraphics[width=0.47\textwidth]{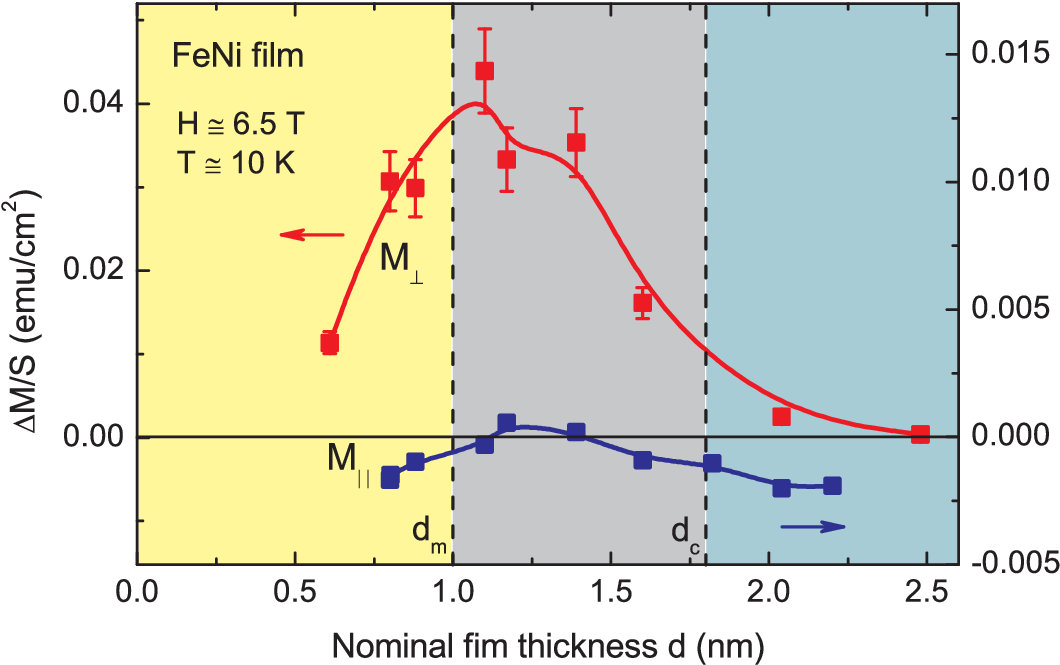}
\caption{Profile of the out-of-plane, $\Delta M_{\perp}/S$, and in-plane, $\Delta M_{\parallel}/S$, magnetization response at the constant magnetic field $H$\,$\simeq$\,6.5 T at $T$\,$\simeq$\,10 K as function of the nominal FeNi film thickness. Here, the thickness $d_m$\,$\simeq$\,1\,nm, close to the SPM\,--\,SFM phase transition, is related to the magnetic percolation point (Sousa et al. 2004). The solid curves are the guides-to-the-eye.}\vspace{-1.2em}
\label{Profile}       
\end{figure}

The observed trends in the magnetization behavior of the studied nanoisland FeNi films, grown onto the Sitall glass substrate, as function of their thickness can be clearly followed from Fig.\,\ref{Profile}, where we present the profile of the out-of-plane and in-plane effective magnetization of the studied FeNi films at the constant external magnetic field $H$\,$\simeq$\,6.5\,T. 

Obviously, the observed out-of-plane and in-plane high-field magnetization response of the inhomogeneous nanoisland FeNi films on the Sitall substrate cannot be associated with conventional magnetism of the permalloy layer. Moreover,   the occurrence of diamagnetic-like response, registered in the in-plane high-field magnetization response, highlights the anomalous behavior of the FeNi films on the Sitall substrate. We noticed that the discovered anomalous anisotropic high-field magnetization behavior has some especially close analogies with the system of gold capped with diamagnetic organic molecules. There, the induced very high specific magnetization, up to several tens of Bohr magnetons per adsorbed molecule, with a very small hysteresis and no saturation up to a field of 1 T, is highly anisotropic, with the highest response along the axis perpendicular to the surface (Carmeli et al. 2003; Vager et al. 2004b).
The observed FM-like unsaturated response is related to charge-transfer (CT) between the organic layer and the metal substrate. It has been proposed that this CT results in unpaired electrons on the 2D organization of the organic molecules on gold, which can provide the basis for the observed magnetism (more details are discussed below).\\

\hspace{-1.0em}{\bf Discussion}.
First, we discuss possible and preferred processes at the interface of the studied system FeNi film/Sitall substrate. Sitall is a glass-ceramic composite material, which is obtained by partial crystallization of a Sitall-forming glass matrix. An analysis of the X-ray diffraction pattern of the Sitall substrate, used in the present experiments, showed that it contains the crystalline TiO$_2$ rutile phase (Kovaleva et al. 2016).
Here, the Sitall-forming glass matrix is represented by TiO$_2$ glass. It is also reported in literature that stimulation of the crystallization process in the starting TiO$_2$ glass may create a conducting surface layer due to partial reduction of Ti$^{4+}$ to Ti$^{3+}$ (Bobkova et al.1994).
In contrast to the highly stoichiometric crystalline Al$_2$O$_3$ substrate, in the Sitall substrate glass phase, the mobile ionic charges of Ti$^{4+}$ and O$^{2-}$ are available. It is known that image interactions due to the charges in the non-metal provide a significant contribution to the metal/non-metal adhesion (Stoneham and Tasker 1985). 
In particular, in the metal/glass adhesion, when a metal is deposited, the image interactions attract the mobile ions of the non-metal glass near to the metal/non-metal interface. Moreover, the wetting conditions are apparently different for the crystalline Al$_2$O$_3$ substrate and the glass-ceramic Sitall (TiO$_2$) substrate. There is a rule that for wetting one needs the optic dielectric constant $\epsilon_{\infty}\gtrsim$\,5.3 (the refractive index $n=\epsilon_{\infty}^{1/2}\gtrsim$\,2.3) for the oxide substrate (Stoneham 1982--83). 
Namely in these systems the catalytic so-called `metal-support interaction' is also appeared to be strong (Stoneham 1982--83).
For example, for Al$_2$O$_3$ substrate, $\epsilon_{\infty}\simeq$\,2.9, which means that liquid metals will not wet it well.
For TiO$_2$, we have $\epsilon_{\infty}\gtrsim$\,6.8\,--\,8.4, which provides good conditions for wetting by liquid metals. 

One may expect that the existence of a conducting surface layer due to partial reduction of Ti$^{4+}$ to Ti$^{3+}$ induced by the crystallization process in the starting TiO$_2$ glass (Bobkova et al.1994) 
and the presence of mobile ionic charges in the Sitall glass may result in high concentration of defects at the FeNi film/Sitall (TiO$_2$) interface. However, we suggest that in the process of FeNi adhesion onto the Sitall (TiO$_2$) substrate, the essential physics is likely to be related to the phenomena disclosed for supported-metal catalysts.  These systems also consist of metallic nanoparticles (usually of group VIII noble metals), dispersed on transition metal oxides, and exhibit unusual properties in catalyst activity in CO and H$_2$ reactions. The nature of the supported-metal interaction, which markedly modifies catalytic and chemical properties of metals, is not yet well understood, however, there is a direct correlation with reducibility of transition metal oxides. Thus, easily reducible transition metal oxides, such as TiO$_2$ and Nb$_2$O$_5$ exhibit strong supported-metal interaction. For example, studies involving supported nickel catalysts show tenfold greater activity for Ni/TiO$_2$ than for Ni/Al$_2$O$_3$ (Vannice and Garten 1979).
The oxide surface reduction, which can be done, for example, by the creation of oxygen vacancy at the TiO$_2$ surface, leads to the formation of Ti$^{3+}$ ions. At the same time, removal of oxygen anions from the surface can provide the conditions for the metal ion and the surface cation be close enough for bonding. 
The process is accompanied by an electron transfer from the cation (such as Ti$^{3+}$ or Nb$^{4+}$) to the metal particle (Tauster et al. 1981).
On an electron transfer from Ti$^{3+}$ ion to the supported metal atom, a Schottky barrier is formed at the interface. Then,  the supported-metal `interaction' may result in the creation of a thin dipolar layer and a Schottky barrier also at the FeNi film/Sitall (TiO$_2$) interface. 

We already noticed that the discovered anomalous anisotropic high-field magnetization behavior of the nano-island FeNi films/Sitall has some especially close analogies with the system of gold capped with diamagnetic organic molecules. There, the induced very high specific magnetization, up to several tens of Bohr magnetons per adsorbed molecule, with a very small hysteresis and no saturation up to a field of 1 T, is highly anisotropic, with the highest response along the axis perpendicular to the surface (Carmeli et al. 2003; Vager et al. 2004b). 
When the organic molecules are self-assembled as monolayers on gold substrate, most of the molecules form chemical bonds with the substrate, and a pseudo-2D layer of dipoles is created. The energetically most effective channel for reducing the created dipolar field is the electron transfer from the gold substrate to the organic molecular layer. 
It is proposed that,  in the system of gold capped with organic molecules, the transferred electrons have large radius around the long molecular axis and are `squeezed' on a two-dimensional network of boundaries between the molecules. It is suggested that this may promote a many body state, where the electrons are paired in spin triplets (Vager and Naaman 2004).
Another major issue for the occurrence of `giant' orbital magnetism, considered for gold capped with organic molecules, is the presence of nanoscale domains formed during the adhesion process (Vager and Naaman 2004, Hernando et al. 2006). 
The orbital angular momentum $l$ for a pair of triplet electrons allows only for odd integers $l=$\,1, 3, ... . In an effective perpendicular magnetic field $H$, composed of an external field and the internal field due to average neighboring magnetization, the energy of a triplet pair of electrons within a circular domain of radius $\xi$ is determined by the following Hamiltonian $\mathcal{H}=T+\mu_Bl_zH$,  where $T$ is the kinetic energy, with the eigenvalues determined by $l$, $E_l(H)=\frac{\hbar^2l^2}{4m_0\xi^2}+\mu_{\rm B}lH=\frac{\hbar^2}{4m_0\xi^2}\left[ (l-\lambda)^2-\lambda^2 \right]$, where $m_0$ is the free electron mass. Here, the lowest energy modes correspond to rotations around the domain axis, with a high angular momentum value $\lambda=-2\frac{\pi\xi^2H}{\Phi_0}=-2\frac{\Phi}{\Phi_0}$, where $\Phi$ is the magnetic flux penetrating the circular domain, and $\Phi_0=\frac{hc}{e}$ is the magnetic flux quantum (Vager and Naaman 2004).
The magnetization of a domain  is estimated as $M=-\mu_{\rm B}\lambda 
\left\langle N \right\rangle$, where $\left\langle N \right\rangle$ corresponds to an average number of triplet pairs in the domain. The above theory provides an explanation for high specific magnetization of the organic molecules self-assembled as monolayers on gold substrate, with the highest response along the axis perpendicular to the surface (Carmeli et al. 2003; Vager et al. 2004b).
Comparing the experimental out-of-plane magnetization with its theoretical value, we can estimate the domain radius $\xi_{\mu}$\,$\approx$\,0.145\,$\mu$m. According to the theory, for this radius, at the field 1\,T, we can have $\left\vert \lambda \right\vert$\,$\simeq$\,1520$H_T\xi_{\mu}^2$\,$\simeq$\,31--33.
 
On the other hand, when a region of the substrate is capped with organic molecules, a potential gradient, which is induced by CT associated with binding, appears at the boundary, extended along the screening length. 
It is suggested that, as a consequence of the contact potential, a radial electric field is induced at the domain boundary, and quasifree electrons can be eventually captured in atomiclike orbitals of large radius $\xi$ at the domain boundary potential step (Hernando et al. 2006).
It is shown that the combination of spin-orbit coupling and contact potential at the large radius $\xi$ domain boundaries can account for the existence of a `giant' orbital moment induced in atomiclike localized states. Moreover, it is argued that the increase of CT reinforces the contact potential and spin-orbit interaction strength, and, consequently, the orbital momentum (Hernando et al. 2006). 

As we have shown above, a thin dipolar layer and a Schottky barrier can be also created at the FeNi film/Sitall (TiO$_2$) interface. Then, we suggest that many free metal electrons may be localized in the edge atomiclike states of large radius $\xi$ at the FeNi nanoisland boundary. Such a cluster, based on a FeNi nanoisland, with many electrons localized in the edge atomiclike states, free metal electrons, and the superspin (SS) magnetic moment, may have lower total energy in a kind of many-body-localized (MBL) state, including electronic, magnetic, and Coulomb contributions. In particular, electrostatic interaction between the localized electrons tends to decrease through the spin alignment, similarly to the effect described by the Hund's rule in atomic orbitals. The detailed mechanism of the ferromagnetic metallic cluster formation in the Kondo-lattice metal, which is accompanied by opening of the pseudogap in the conduction band of the itinerant charge carriers and development of the low- and high-spin intersite electronic transitions, is discussed by Kovaleva (Kovaleva et al. 2012, 2004, 2007, 2010). Further, the spin alignment of the electrons localized in the edge atomiclike states implies the alignment with the orbital moment  through the spin-orbit coupling. In general, we expect to find a very broad distribution of such clusters in the inhomogeneous nanoisland FeNi films of different nominal thickness. Then, it is reasonable to suggest that a variety of the low-temperature `exotic' magnetic phases, which can be found in the inhomogeneous FeNi nanoisland FeNi films, is associated with the existence of such magnetic clusters, as described above. In the thinnest FeNi films under the study, the system may exist in the `orbital glass' state, similar to that described by Kusmartsev (Kusmartsev 1992). 
In the sufficiently dense FeNi films, FM-like coupling between large orbital moments of the nanoislands can occur. FM-like coupling between large orbital moments of the nanoislands can contribute to the SFM behavior found in self-assembled quasi-2D metallic magnetic FeNi nanoislands. In the SFM phase, the clusters can merge into ordered stripy microdomains. 
As it was shown in our previous study (Kovaleva et al. 2012), 
at high temperatures, this state can survive until the critical temperature, associated with delocalization of the electrons from the edge atomiclike states at the boundary of the nanoisland. 

The observed anomalous out-of-plane paramagnetic-like magnetization response associated with a large orbital moment of localized electrons can lead to an additional contribution to magnetization induced in the used Sitall substrate with the SPM impurities. In fact, the SPM Sitall substrate can reenforce the observed out-of-plane paramagnetic- and diamagnetic-like effects. However, the out-of-plane magnetization and in-plane demagnetization of the SPM Sitall substrate prevent accurate evaluation of the effective paramagnetic- and diamagnetic-like susceptibilities.      

From Fig.\,\ref{Profile}, where we present the profile of the out-of-plane and in-plane effective magnetization of the studied inhomogeneous nanoisland FeNi films at the constant external magnetic field $H$\,$\simeq$\,6.5\,T, one notices that the anomalous in-plane high-field diamagnetic-like response was observed for the FeNi films/Sitall where the out-of-plane SFM phase is suppressed.  The diamag-netic-like response, competing with the FM-like response for the FeNi film samples with the nominal thickness of 0.8, 1.1 and 1.6\,nm, becomes clearly recognizable above 1.5\,--\,2\,T at large $H$ (see Fig.\,\ref{DeltaM}(b)). Moreover, seemingly, the out-of-plane SFM coupling is almost suppressed near the percolation threshold, and, for the FeNi film with the thickness 2.0\,nm, above the FeNi film percolation threshold due to the nanoisland coalescence at the $d_c$\,$\simeq$\,1.8 nm, the in-plane diamagnetic-like magnetization response was observed for the whole  range of studied dc magnetic fields (see Fig.\,\ref{DeltaM}(b)). 

We suggest that the discovered in-plane diamagnetic-like response can arise from the low-temperature `exotic' magnetic phase, associated with the existence of such magnetic FeNi clusters, as described above. Thus, it is established that in the systems with an effective perpendicular anisotropy, there is a tendency to inhomogeneous distribution of magnetic moments in the form of supervortices (SVs) (Stupakov et al. 2016; Dzian et al. 2013). 
Then, the observed behavior can be discussed in terms of the non-superconducting diamagnetic state related to the inhomogeneous distribution of the toroidal moment density (Volkov et al. 1984). 
In particular, inhomogeneous toroidal moment density can be generated in magnetic materials with inhomogeneous magnetization, where the formation of ordered vortex-like states is favored. Here each magnetic vortex possesses toroidal moment directed along the vortex line. In the parameter range corresponding to the system electron percolation toroidal inhomogeneity can create orbital magnetic filed  (Gorbatsevich 1989), which can result in anomalously strong diamagnetic response (Ginzburg et al. 1984).
This may be well consistent with the in-plane diamagnetic-like response found in the present study for the inhomogeneous nanoisland FeNi films with the nominal thickness around the percolation threshold at the $d_c$\,$\simeq$\,1.8\,nm.  The observed diamag-netic-like response can be explained if one assumes the existence of rigid vortex-like magnetic structure with inhomogeneously distributed vortices. Inhomogeneous vortex distribution acts on percolating electrons as an effective magnetic filed which forms spontaneous current contours exhibiting Larmour's precession in an external magnetic field  (Gorbatsevich 1989).                

In relation to this, we would like to mention that observation of the anomalous almost `ideal diamagnetic and paramagnetic' response, recorded at high magnetic fields, was reported for epitaxial layers of CuCl on Si substrates by Mattes and Foiles (Mattes and Foiles 1985). 
There, a possible mechanism of the anomalous high-field magnetic response was attributed to interface magnetism, related to the energy band heterostructure at the interface, and given in a framework of the excitonic high-temperature superconductivity mechanism. However, we would like to note that, in this earlier article, the role of a domain structure for the occurrence of the anomalously large diamagnetic response was also discussed, in lines with the theory (Ginzburg et al. 1984).
\\ 

\hspace{-1.0em}{\bf Conclusions}.
By using SQUID magnetometry we investigated anisotropic magnetization response of the inhomogeneous nanoisland FeNi 
films grown by rf sputtering deposition onto the Sitall (TiO$_2$) glass substrates. The obtained magnetization response is clearly beyond conventional spin-magnetism of Fe$_{21}$Ni$_{79}$ permalloy. We have found that for the nanoisland FeNi films with the nominal thickness below the physical percolation threshold, varying from 0.6 to 1.6 nm, the out-of-plane magnetization response is not saturated in the range of investigated magnetic fields (0\,T\,$\lesssim$\,$H$\,$\lesssim$\,7\,T) and exhibits paramagnetic-like magnetization response. The magnetization response reveals an escalating slope with increasing the nominal film thickness from 0.6 to 1.1 nm. The maximum slope is obtained for the 1.1\,--\,1.17\,nm FeNi film thicknesses. With further increasing the film thickness, the slope is decreasing, and the anomalous out-of-plane paramagnetic-like response vanishes on approaching the FeNi film percolation threshold at the $d_c$\,$\simeq$\,1.8\,nm. In addition, the extracted high-field magnetization response exhibits remarkable anisotropy. Interesting, we have found that when the anomalous out-of-plane paramagnetic-like magnetization response becomes suppressed in the nanoisland FeNi films, the ano-malous in-plane diamagnetic-like response becomes evident above 1.5\,--\,2\,T at high magnetic fields. Moreover, we have discovered that for the FeNi film with the thickness 2.0\,nm, above the FeNi film 
percolation threshold, the anomalously large in-plane diamagnetic-like magnetization response was observed for a whole range of the studied dc magnetic fields 0\,T\,$\lesssim$\,$H$\,$\lesssim$\,7\,T. 

The origin of the `exotic' magnetic phases, discovered in the inhomogeneous FeNi nanoilsand films/Sitall (TiO$_2$) and identified by their anomalous paramagnetic-  and diamagnetic-like response, is discussed. The found high-field magnetic properties of the inhomogeneous nanoiland FeNi films can be utilized, for example, in high magnetic field sensing devices. \\ 

The experiments were performed in the Materials Growth and Measurement Laboratory MGML (http://mgml.eu). This work was supported by the grant N17-72-20030 of the RSF (Russian Science Foundation). \\

\hspace{-1.7em}{\bf References}\\

\hspace{-1.7em}Bobkova NM, Barantseva SE, Gailevich SA (1994) Composite materials based on sitall glass and technical aluminum and titanium oxides. Glass and Ceramics 51(11-12):335-339. 

\hspace{-1.7em}Boltaev AP, Pudonin FA, Sherstnev IA, Egorov DA (2017) Detection of the metal-insulator transition in disordered systems of magnetic nanoislands, JETP 125(3):465-468. 
https://doi.org/10.1134/S1063776117080027  

\hspace{-1.7em}Carmeli I, Leitus G, Naaman R, Reich S, Vager Z (2003) Magnetism induced by the organization of self-assembled monolayers. J Chem Phys 118(23):10372-10375. https://doi.org/10.1063/1.1580800

\hspace{-1.7em}Cowburn RP, Koltsov DK, Adeyeye AO, Welland ME, Tricker DM (1999) Single-domain circular nanomagnets. Phys Rev Lett 83(5):1042-1045. https://doi.org/10.1103/PhysRevLett.83.1042

\hspace{-1.7em}Dzian SA, Galkin AY, Ivanov BA, Kireev VE, Muravyov VM (2013) Vortex ground state for small arrays of magnetic particles with dipole coupling. Phys Rev B 87(18):184404. https://doi.org/10.1103/PhysRevB.87.184404 

\hspace{-1.7em}Elmen GW (1935) Magnetic alloys of Iron, Nickel, and Cobalt, Electrical Engineering. https://doi.org/10.1002/j.1538-7305.1929.tb04428.x

\hspace{-1.7em}Frey NA, Sun S (2011) Magnetic Nanoparticle for Information Storage Applications. Inorganic Nanoparticles: Synthesis, Applications, and Perspectives. 
Edited by Altavilla C, Ciliberto E. 
CRC Press. Taylor \& Francis Group.

\hspace{-1.7em}Ganichev SD, Ivchenko EL, Bel'kov VV, Tarasenko SA, Sollinger M, Weiss D, Wegscheider W, Prettl W (2002) Spin-galvanic effect. Nature 417:153-156. https://doi.org/10.1038/417153a

\hspace{-1.7em}Ginzburg VL, Gorbatsevich AA, Kopaev YuV, Volkov BA (1984) On the problem of superdiamagnetism. Solid State Commun 50(4):339-343. https://doi.org/10.1016/0038-1098(84)90381-8

\hspace{-1.7em}Gorbatsevich AA (1989) Quasimomentum spectral asymmetry and the anomalous magnetic properties of an orbital antiferromagnet. Sov Phys JETP 68(4): 847-856.

\hspace{-1.7em}Hernando A, Crespo P, Garc\'ia MA (2006) Origin of orbital ferromagnetism and giant magnetic anisotropy at the nanoscale. Phys Rev Lett 96(5):057206. https://doi.org/10.1103/PhysRevLett.96.057206

\hspace{-1.7em}Kern PR, Silva OE, Siqueira JV, Della Pace RD, Rigue JN, Carara M (2016) A study on the thickness dependence of static and dynamic magnetic
properties of Ni$_{81}$Fe$_{19}$ thin films. JMMM 419:456-–463. https://doi.org/10.1016/j.jmmm.2016.06.061

\hspace{-1.7em}Kleemann W, Petracic O, Binek Ch, Kakazei GN, Pogorelov YuG, Sousa JB, Cardoso S, Freitas PP (2001) Interacting ferromagnetic nanoparticles in discontinuous Co$_{80}$Fe$_{20}$/Al$_2$O$_3$ multilayers: From superspin glass to reentrant superferromagnetism. Phys Rev B 63(13):134423. 
https://doi.org/10.1103/PhysRevB.63.134423

\hspace{-1.7em}Kondratyev VN, Lutz HO (1998) Shell effect in exchange coupling of transition metal dots and their arrays. Phys Rev Lett 81(20):4508-4511. https://doi.org/10.1103/PhysRevLett.81.4508

\hspace{-1.7em}Kovaleva NN, Boris AV, Bernhard C, Kulakov A, Pimenov\,A, Balbashov AM, Khaliullin G, Keimer B (2004) Spin-controlled Mott-Hubbard bands in LaMnO$_3$ probed by optical ellipsometry. Phys Rev Lett 93(14):147204. https://doi.org/10.1103/PhysRevLett.93.147204

\hspace{-1.7em}Kovaleva NN, Boris AV, Yordanov P, Maljuk A, E. Br\"ucher, Strempfer J, Konuma M, Zegkinoglou I, Bernhard C, Stoneham AM, Keimer B (2007) Optical response of ferromagnetic YTiO$_3$ studied by spectral ellipsometry. Phys Rev B 76(15):155125. https://doi.org/10.1103/PhysRevB.76.155125  

\hspace{-1.7em}Kovaleva NN, Ole\'s AM, Balbashov AM, Maljuk A, Argyriou DN, Khaliullin G, Keimer B (2010) Low-energy Mott-Hubbard excitations in LaMnO$_3$ probed by optical ellipsometry. Phys Rev B 81(23):235130. https://doi.org/10.1103/PhysRevB.81.235130

\hspace{-1.7em}Kovaleva NN, Kugel KI, Bazhenov AV, Fursova TN, W. L\"oser W, Xu Y, G. Behr G, Kusmartsev FV (2012) Formation of metallic magnetic clusters in a Kondo-lattice metall: Evidence from an optical study. Sci Rep 2, article 890:1-7. https://doi.org/10.1038/srep00890

\hspace{-1.7em}Kovaleva NN, Chvostova D, Bagdinov AV, Petrova MG, Demikhov EI, Pudonin FA, Dejneka A (2015) Interplay of electron correlations and localization in disordered $\beta$-tantalum films: Evidence from dc transport and spectroscopic ellipsometry study. Appl Phys Lett 106:051907. http://doi.org/10.1063/1.4907862

\hspace{-1.7em}Kusmartsev FV (1992) Orbital glass. Phys Lett A 169(1-2):108-114. https://doi.org/10.1016/0375-9601(92)90815-4 

\hspace{-1.7em}Mattes L, Foiles CL (1985) Large diamagnetism and paramagnetism with a CuCl:Si interface. Physica B 135(1-3):139-147. \\https://doi.org/10.1016/0378-4363(85)90454-1 

\hspace{-1.7em}McNally A (2013) Magnetic sensors: Nanoparticles detect infection. Nature Nanotech 8:315-316. https://doi.org/10.1038/nnano.2013.76

\hspace{-1.7em}Miu D, Jinga SI, Vasile BS, Miu L (2015) Out of plane superferromagnetic behavior of quasi two-dimensional Fe/Al$_2$O$_3$ multilayer nanocomposites. Appl Phys Lett 117, 074303. https://doi.org/10.1063/1.4908219

\hspace{-1.7em}Mody VV, Cox A, Shah S, Singh A, Bevins W, Parihar H (2014) Magnetic nanoparticle drug delivery systems. Appl Nanosci 4(4): 385-392. https://doi.org/10.1007/s13204-013-0216-y

\hspace{-1.7em}Oja R, Tyunina M, Yao L, Pinomaa T, Kocourek T, Dejneka A, Stupakov O, Jelinek M, Trepakov VV, van Dijken S, Nieminen RM (2012) Ferromagnetic interface between nonmagnetic perovskites. Phys Rev Lett 109(12):127207. http://doi.org/10.1103/PhysRevLett.109.127207

\hspace{-1.7em}Rancourt DG, Daniels JM(1984) Influence of unequal magnetization direction probabilities on the M\"ossbauer spectra of superparamagnetic particles. Phys Rev B 29(5): 2410-2414. https://doi.org/10.1103/PhysRevB.29.2410 

\hspace{-1.7em}Reiss G, Hutten A (2005) Magnetic nanoparticles: Applications beyond data storage. Nature Mater 4:725-726. https://doi.org/10.1038/nmat1494

\hspace{-1.7em}Rumpf K, Granitzer P, Krenn H (2008) Beyond spin-magnetism of magnetic nanowires in porous silicon. J Phys: Condens Matter 20(45):454221. https://doi.org/10.1088/0953-8984/20/45/454221

\hspace{-1.7em}Rumpf K, Granitzer P, Poelt P (2010) Non-saturating magnetic behaviour of a ferromagnetic semiconductor/metal nanocomposite. JMMM 322(9-12):1283-1285. https://doi.org/10.1016/j.jmmm.2009.04.075 

\hspace{-1.7em}Russier V (2001) Calculated magnetic properties of two-dimensional arrays of nanoparticles at vanishing temperature. J Appl Phys 89(2):1287-1294. https://doi.org/10.1063/1.1333034 

\hspace{-1.7em}Schneider M, Hoffmann H (1999) Magnetization loops of submicron ferromagnetic permalloy dot arrays. J Appl Phys 86(8):4539-4543. https://doi.org/10.1063/1.371399

\hspace{-1.7em}Sherstnev IA (2014) Electronic transport and magnetic structure of nanoisland ferromagnetic materials systems. Ph.D. thesis.
http://www.lebedev.ru/file/910

\hspace{-1.7em}Sousa JB, Santos JAM, Silva RFA, Teixeira JM, Ventura J, Ara\'ujo JP, Freitas PP, Cardoso S, Pogorelov YuG, Kakazei GN, Snoeck (2004) 
Peculiar magnetic and electrical properties near structural percolation in metal-insulator granular layers. J Appl Phys 96(7):3861-3864. https://doi.org/10.1063/1.1786651

\hspace{-1.7em}Stoneham AM (1982-83) Systematics of metal-insulator interfacial energies: A new rule for wetting and strong catalyst-support interactions. Appl Surf Sci 14:249-259. 

\hspace{-1.7em}Stoneham AM, Tasker PW (1985) Metal-non-metal and other interfaces: the role of image interactions. J Phys C: Solid State Phys 18(19):L543-L548.   

\hspace{-1.5em}Stupakov A, Bagdinov AV, Prokhorov VV, Bagdinova AN, Demikhov EI, Dejneka A, Kugel KI, Gorbatsevich AA, Pudonin FA, Kovaleva NN (2016) Out-of-plane and in-plane magnetization behavior of dipolar
interacting FeNi nanoislands around the percolation threshold. J Nanomater 2016, Article ID 3190260, 9 pages. http://doi.org/10.1155/2016/3190260

\hspace{-1.7em}Sugawara A, Hembree GG, Scheinfein MR (1997) Self-organized Fe nanowire arrays prepared by shadow deposition on NaCl(110) templates. J Appl Phys 70(8):1043-1045. https://doi.org/10.1063/1.118437 

\hspace{-1.7em}Sun S, Murray CB (1999) Synthesis of monodisperse cobalt nanocrystals and their assembly into magnetic superlattices. J Appl Phys 85(8):4325-4390. https://doi.org/10.1063/1.370357 

\hspace{-1.7em}Tauster SJ, Fung SC, Baker RTK, Horsley JA (1981) Strong interactions in suported-metal catalysts. Science 211(4487):1121-1125. https://doi.org/10.1126/science.211.4487.1121

\hspace{-1.7em}Vager Z, Naaman R (2004) Bosons as the origin for giant magnetic properties of organic monolayers. Phys Rev Lett 92(8):087205. https://doi.org/10.1103/PhysRevLett.92.087205

\hspace{-1.7em}Vager Z, Carmeli I, Leitus G, Reich S, Naaman R (2004) Surprising electronic-magnetic properties of closed packed organized organic layers. J Phys Chem Solids 65(4):713-717. https://doi.org/10.1016/j.jpcs.2003.11.031

\hspace{-1.7em}Vannice MA, Garten RL (1979) Metal-support effects on the activity and selectivity of Ni catalysts in CO-H$_2$ synthesis reactions. J Catal 56(2):236-248.


\hspace{-1.7em}Volkov BA, Gorbatsevich AA, Kopaev YuV (1984) Diamagnetic anomalies in systems with spontaneous toroidal current. Sov Phys JETP 59(5):1087-1098.

\end{document}